\documentclass{article}
\usepackage{spconf,amsmath,graphicx,hyperref}
\usepackage{booktabs}
\usepackage{multirow}
\title{FROM PDF TO EVIDENCE: STRUCTURE-AWARE RETRIEVAL \\ FOR CLINICAL PRACTICE GUIDELINES}

\name{Xingyu Lin, Dehui Du}
\address{Software Engineering Institute, East China Normal University \\ Shanghai, China \\ xingyulin@stu.ecnu.edu.cn, dhdu@sei.ecnu.edu.cn}

\begin{document}
\frenchspacing
\ninept
\maketitle

\begin{abstract}
Guideline documents are published as unstructured PDFs whose evidence is locked in
visual structures---tables, flowcharts, and graded recommendations---that standard
retrieval pipelines flatten into fixed-size text chunks. We cast evidence access as a
document image analysis problem: parse each page image into typed structural elements,
then retrieve structure-aware evidence units that follow the document's own layout
(sections, table rows, flowchart paths, graded recommendations), each keeping its
structural context so a result points to a specific element rather than a page. On 26
clinical practice guidelines from 9 sources (3{,}619 pages, Chinese and English) with 199
evidence queries, structure-aware units rank the gold element first under BM25, dense,
and hybrid retrieval (hybrid Element Hit@1 of 0.382), with a significant element-level
ranking gain over per-element OCR text (MRR$_e$ $+0.107$, $p{=}0.002$; the Hit@5 gain is
directional, $p{=}0.17$), while matching page-level recall (Page Hit@5 0.879 vs.\ 0.889,
$p{=}0.75$) at 3.8$\times$ less context and clearly outperforming a ColPali visual-RAG
baseline (PH@5 0.497).
\end{abstract}


\section{Introduction}
\label{sec:intro}

Clinical practice guidelines (CPGs) from NCCN, WHO, and CSCO arrive as unstructured
PDFs---born-digital or scanned page images---whose evidence is locked in visual
structures: treatment algorithms are flowcharts with branching paths, dosing lives in
tables with header-value relationships, and evidence grades (NCCN Category~1, CSCO
Level~I) sit in recommendation paragraphs. Exposing this evidence to natural-language
queries is thus a document image analysis problem as much as a retrieval one. Current
pipelines treat these documents as flat text---either fixed-size
chunks~\cite{santos2026pdfragready} or whole pages~\cite{faysse2024colpali}. Chunking
cuts tables apart, so a dose value lands in a different chunk from the column header
that gives it meaning; page-level retrieval finds the right page but hands the LLM
thousands of tokens for a few lines of evidence.

Structure preservation in retrieval has drawn recent attention. SPIRE~\cite{rainey2026spire}
shows tree-structured extraction improves evidence retrieval from HTML, SEM-RAG~\cite{yang2026semrag}
compiles layout-aware graphs for telecommunication standards, and clinical
systems such as Guideline2Graph~\cite{kilic2026guideline2graph} convert guidelines into
decision graphs; MedDM~\cite{li2023meddm} and Flow2MDT~\cite{zhai2025flow2mdt} build
decision trees from flowchart images, and GGPONC~\cite{borchert2020ggponc} demonstrates
the value of guideline-derived corpora with evidence-level metadata. On the parsing side,
layout benchmarks (PubLayNet~\cite{zhong2019publaynet}, DocLayNet~\cite{pfitzmann2022doclaynet},
OmniDocBench~\cite{ouyang2025omnidocbench}) and criteria for text extraction from
layout-based formats~\cite{bast2017pdf} matured quickly, hierarchical structure
recovery~\cite{wang2024detect} and table reconstruction~\cite{nassar2022tableformer}
advanced in parallel, but none of these benchmarks contains clinical
guidelines, and RAG-readiness evaluations on administrative
documents~\cite{santos2026pdfragready} do not transfer. What remains unmeasured is how structure preservation at the parsing stage
affects evidence retrieval across the formats CPGs actually come in. Recent end-to-end
parsers (DeepSeek-OCR~\cite{wei2025deepseekocr}, dots.ocr~\cite{li2025dotsocr},
Unlimited-OCR~\cite{yin2026unlimitedocr}) now emit typed elements with bounding boxes from
page images, and visual retrievers such as ColPali~\cite{faysse2024colpali} retrieve page
images directly---but none of these exposes element-level evidence grounded in the
document's own table, flowchart, and recommendation structure. Clinical RAG
systems~\cite{kresevic2024hepatological,lewis2025niceguideline,xiong2024medrag,zakka2024almanac,jeong2024selfbiorag,li2025clinicbot,li2025regionrag}
depend on this evidence but assume the parsing stage has already produced clean, structured
output; cpgQA~\cite{mahbub2023cpgqa} excludes tables and figures altogether, and
CPGBench~\cite{tan2026cpgbench} evaluates conversational adherence rather than parsing
fidelity. Hsu et al.~\cite{hsu2025extracting} extract pharmacogenomic recommendations but
start from already-curated text, not the source PDFs.

We propose \textbf{structure-aware evidence units}: four unit types that mirror the
document's structure---(1)~\emph{section units}, a heading and its following paragraphs;
(2)~\emph{table row units}, one per row with column headers attached; (3)~\emph{flowchart
path units}, one per directed edge of a clinical algorithm; and (4)~\emph{recommendation
units}, paragraphs carrying evidence-grade markers. Each unit keeps its structural context
(table headers, flowchart titles, section headings), so a retrieval result points to a
specific element. We evaluate on 26 CPGs from 9 sources and 16 specialties, Chinese and
English, born-digital and scanned, with 199 evidence queries. Using BM25, dense
(Qwen3-Embedding-8B), and hybrid retrieval, we compare against flat chunking,
heading-aware chunking, page-level retrieval, per-element OCR text, and a ColPali
visual-RAG baseline. Structure-aware units rank the gold element first under every retriever, at a fraction of the context cost of page-level retrieval---and
outperform the ColPali visual baseline by a wide margin on page-level recall (PH@5
0.879 vs.\ 0.497).

\begin{figure}[t]
\centerline{\includegraphics[width=\columnwidth]{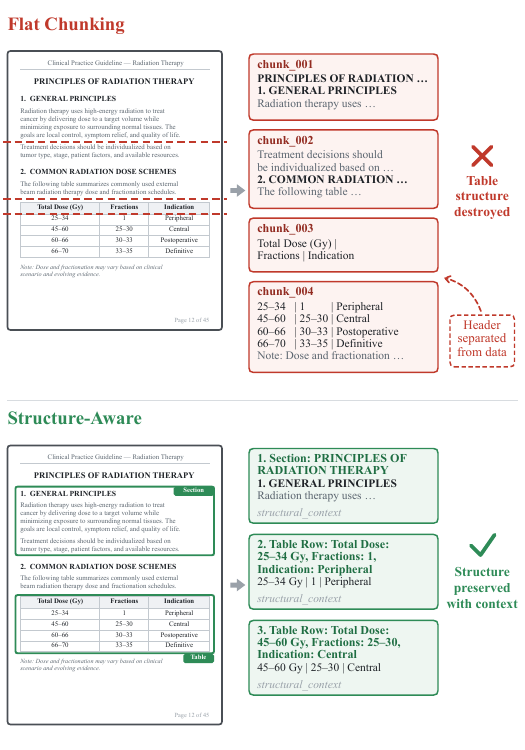}}
\caption{Flat chunking (top) splits table headers from data; a structure-aware unit
(bottom) keeps each row attached to its headers and table context.}
\label{fig:comparison}
\end{figure}

\section{Method}
\label{sec:method}

\subsection{Overview}
\label{ssec:overview}

\begin{figure}[t]
\centerline{\includegraphics[width=\columnwidth]{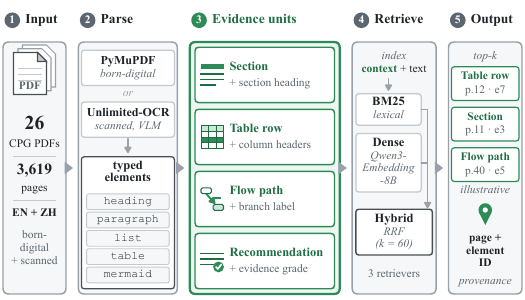}}
\caption{Pipeline overview. Pages are parsed into typed elements, assembled into four
structure-aware evidence unit types (Section / Table row / Flow path / Recommendation)
that keep their structural context, and retrieved with BM25, dense, or hybrid fusion,
yielding evidence with page- and element-level provenance (output examples illustrative).}
\label{fig:overview}
\end{figure}

Fig.~\ref{fig:overview} illustrates the pipeline. Given a set of CPG PDFs, we first extract page-level structured content with an OCR/VLM
pipeline, producing typed elements (headings, paragraphs, tables, flowcharts) with unique
identifiers. We then construct four types of structure-aware evidence units, each
preserving its structural context. At query time we retrieve with BM25, dense embedding, or
hybrid fusion, returning fine-grained evidence units that point back to specific page
elements.

\subsection{Parsing into typed elements}
\label{ssec:parsing}

We support both born-digital PDFs (native text layers, extracted via PyMuPDF and classified
by layout patterns) and scanned PDFs (image-only), for which we use
Unlimited-OCR~\cite{yin2026unlimitedocr}, an end-to-end vision-language model that outputs
element-level detections with bounding boxes and type labels. Both routes normalize into a
shared schema: HTML tables become header and row lists, and flowchart regions are
transcribed into mermaid graphs by a vision-language model. Each element receives a unique
identifier $e_{ij}$ (page $i$, element $j$) and a type label; the types that drive unit
construction are \texttt{heading}, \texttt{paragraph}, \texttt{list}, \texttt{table}, and
\texttt{mermaid}.

\subsection{Structure-aware evidence unit construction}
\label{ssec:units}

We traverse each page's typed elements in reading order, tracking the most recent heading
as the running section context, and emit one unit per structural element in a single linear
pass. \textbf{Section units} concatenate a heading $h$ with the paragraphs that follow it
up to the next heading, with context $c=\texttt{"Section: "}\oplus h$. \textbf{Table row
units} create one unit per row $r_j=(r_{j1},\dots,r_{jm})$ of a table with headers
$[h_1,\dots,h_m]$, as $(h_1\texttt{: }r_{j1}\texttt{, }\dots\texttt{, }h_m\texttt{: }r_{jm})$
with the table caption and headers as context. \textbf{Flowchart path units} parse directed
edges $A\!\to\!B$ (with branch labels) from mermaid-encoded algorithms, one unit per edge.
\textbf{Recommendation units} are paragraphs matching recommendation language or evidence
grades (e.g., ``Category [1-3]'', CSCO Level-I/II/III, class 1A/2B), inheriting their
parent section as context.

\subsection{Retrieval with structural context}
\label{ssec:retrieval}

Each unit $u=(\text{text},c)$ pairs content text with a structural context string. BM25~\cite{robertson2009bm25}
indexes $c\oplus\text{text}$, so structural metadata (table headers, section titles)
contributes to matching; dense retrieval encodes $c\oplus\text{text}$ with
Qwen3-Embedding-8B~\cite{qwen2025embedding}. Hybrid retrieval applies reciprocal rank
fusion ($k{=}60$) over BM25 and dense results.

\section{Experimental Setup}
\label{sec:setup}

\subsection{Guideline corpus}
\label{ssec:corpus}

We assemble 26 CPGs from 9 source collections (NCCN, CSCO, CMA, CDC, NICE, NHC, ChinaCDC,
and the English and Chinese WHO collections), 16 specialties, 3{,}619 pages in English (18
documents) and Chinese (8), both born-digital and scanned. Each page is parsed by
Unlimited-OCR~\cite{yin2026unlimitedocr} into 53{,}595 typed elements; flowchart regions in
the algorithm-heavy NCCN guideline are further processed by Qwen3.6-27B into mermaid
decision paths. Unit construction yields 21{,}662 structure-aware units.

\begin{table}[t]
\caption{Guideline corpus and query-set statistics.}
\label{tab:corpus}
\centering
\footnotesize
\begin{tabular*}{\columnwidth}{@{\extracolsep{\fill}}lcc@{}}
\toprule
\textbf{Dimension} & \textbf{Count} & \textbf{Breakdown} \\
\midrule
\multicolumn{3}{@{}l}{\textit{Corpus}} \\
Guidelines & 26 & 9 sources \\
Pages & 3{,}619 & EN 18 / ZH 8 docs \\
Specialties & 16 & born-digital + scanned \\
OCR elements & 53{,}595 & 5 element types \\
Struct-aware units & 21{,}662 & 4 unit types \\
\midrule
\multicolumn{3}{@{}l}{\textit{Queries}} \\
Annotated queries & 199 & all 26 guidelines \\
\quad by type & --- & table 83 / text 101 / flow 15 \\
\quad by language & --- & EN 114 / ZH 85 \\
\quad manual review & 40 & 20\% random sample \\
\bottomrule
\end{tabular*}
\end{table}

\subsection{Evidence queries}
\label{ssec:queries}

We construct 199 evidence-seeking queries with gold annotations spanning all 26 guidelines.
Each query is a natural-language clinical question (e.g., ``What spinal cord dose
constraint is recommended for 1-fraction SABR?''); gold annotations specify the source PDF,
page, element ID, and evidence span. Queries are stratified by evidence type: 83
table-dependent, 101 text-dependent, 15 flowchart-dependent (114 English, 85 Chinese). To
limit circularity, queries were generated from high-information evidence units but phrased
as standalone clinical questions whose wording is independent of any unit's surface text,
and every retrieval method is evaluated against the \emph{same} gold element IDs---so any
representation bias is shared across all compared methods. An automated check verified
each gold element ID exists and matches the query type, followed by manual review of a
random 20\% sample (40 queries) for clinical naturalness and answer correctness.

\subsection{Methods and metrics}
\label{ssec:methods}

We compare five unit-construction methods: \textbf{flat chunk} (500 chars, 50 overlap),
\textbf{heading-aware chunk}, \textbf{page-level} (one unit per page), \textbf{OCR element}
(each element alone, no context), and \textbf{structure-aware} (ours). Each is run with
BM25, dense, and hybrid retrieval. We additionally compare against
\textbf{ColPali}~\cite{faysse2024colpali}, a visual-RAG baseline that embeds page images
and retrieves by late interaction, evaluated at the page level. Metrics: \textbf{Page
Hit@$k$} (a top-$k$ result on the gold page), \textbf{Element Hit@$k$} (a top-$k$ result
matching the gold element ID; N/A for methods without element addressability),
\textbf{MRR}, and \textbf{context cost} (mean characters in top-5).

\section{Results}
\label{sec:results}

\subsection{Main results}
\label{ssec:main}

Table~\ref{tab:main} presents the main retrieval results. Structure-aware units achieve the
highest element-level localization under all three retrievers. With hybrid retrieval they reach EH@1 of 0.382 and MRR$_e$ of 0.445 at 3{,}222 characters of context. Page-level
retrieval has the highest PH@5 (0.889 vs.\ 0.879) and the best MRR$_p$ under BM25, but it
cannot point below the page and consumes 3.8$\times$ more context (12{,}111 vs.\ 3{,}222
characters). Under the dense retriever, structure-aware units lead page-level recall
outright (PH@1 0.683, PH@5 0.909), because the attached context strings are short prose the
embedding model can match even when wording differs.

\begin{table}[t]
\caption{Main retrieval results. EH@$k$ is N/A for methods without element-level
addressability. Best per retriever in \textbf{bold}.}
\label{tab:main}
\centering
\footnotesize
\setlength{\tabcolsep}{3.2pt}
\begin{tabular*}{\columnwidth}{@{\extracolsep{\fill}}llccccc@{}}
\toprule
\textbf{Retr.} & \textbf{Method} & \textbf{PH@1} & \textbf{PH@5} & \textbf{EH@1} & \textbf{EH@5} & \textbf{Ctx} \\
\midrule
\multirow{5}{*}{BM25}
& Flat chunk      & 0.387 & 0.724 & N/A & N/A & 2{,}495 \\
& Heading chunk   & 0.502 & 0.769 & N/A & N/A & 2{,}080 \\
& Page-level      & 0.558 & \textbf{0.874} & N/A & N/A & 12{,}117 \\
& OCR element     & 0.362 & 0.719 & 0.186 & 0.432 & 2{,}314 \\
& \textbf{Struct-aware} & \textbf{0.568} & 0.809 & \textbf{0.327} & \textbf{0.497} & 3{,}198 \\
\midrule
\multirow{5}{*}{Dense}
& Flat chunk      & 0.513 & 0.789 & N/A & N/A & 2{,}496 \\
& Heading chunk   & 0.598 & 0.859 & N/A & N/A & 2{,}079 \\
& Page-level      & 0.513 & 0.849 & N/A & N/A & 12{,}999 \\
& OCR element     & 0.533 & 0.794 & 0.322 & 0.518 & 1{,}504 \\
& \textbf{Struct-aware} & \textbf{0.683} & \textbf{0.909} & \textbf{0.402} & \textbf{0.573} & 2{,}957 \\
\midrule
\multirow{5}{*}{Hybr.}
& Flat chunk      & 0.457 & 0.804 & N/A & N/A & 2{,}499 \\
& Heading chunk   & 0.593 & 0.874 & N/A & N/A & 2{,}121 \\
& Page-level      & 0.623 & \textbf{0.889} & N/A & N/A & 12{,}111 \\
& OCR element     & 0.447 & 0.789 & 0.241 & 0.487 & 2{,}105 \\
& \textbf{Struct-aware} & \textbf{0.668} & 0.879 & \textbf{0.382} & \textbf{0.543} & 3{,}222 \\
\midrule
\multicolumn{2}{l}{ColPali (visual, page-level)} & 0.332 & 0.497 & N/A & N/A & --- \\
\bottomrule
\end{tabular*}
\end{table}

The ColPali visual-RAG baseline retrieves at the page level, reaching PH@1 of 0.332, PH@5 of 0.497, and MRR$_p$ of 0.410. Structure-aware units under hybrid retrieval clearly outperform it on page-level recall (PH@5 0.879 vs.\ 0.497), while uniquely providing element-level addressability (EH@5 0.543) and using 3.8$\times$ less context than page-level text retrieval.

\subsection{Statistical significance}
\label{ssec:significance}

For each pair (structure-aware vs.\ baseline) we compute a paired bootstrap two-sided
$p$-value (10{,}000 resamples) and, for binary hit metrics, an exact McNemar test. Under
hybrid retrieval, structure-aware units significantly outperform flat chunking on PH@5
($\Delta{=}+0.075$, $p{=}0.004$) and MRR$_p$ ($\Delta{=}+0.156$, $p{<}0.001$). Against the
per-element OCR baseline they achieve significantly higher PH@5 ($\Delta{=}+0.090$,
$p{=}0.003$), MRR$_p$ ($\Delta{=}+0.167$, $p{<}0.001$), and MRR$_e$ ($\Delta{=}+0.107$,
$p{=}0.002$); the element-level hit-rate advantage (EH@5 $+0.055$) is directional but not
significant ($p{=}0.17$). Against heading-aware and page-level baselines, PH@5 differences
are small and not significant ($p{=}0.91$ and $p{=}0.75$), consistent with our claim that
structure-aware retrieval matches page-level recall while providing element-level
addressability at lower context cost.

\subsection{Analysis by query type and language}
\label{ssec:analysis}

At the page level the methods are close across query types; the clearer separation is at
the element level (Table~\ref{tab:pertype}). The advantage of structure-aware units
concentrates on table-dependent queries, where ElementHit@5 rises to 0.747 against 0.506
for the OCR baseline---table-row units keep each row individually addressable with its
headers. On text-dependent queries the ordering flips (0.436 vs.\ 0.535): section units
bundle several paragraphs, diluting the exact gold element. Flowchart-dependent hits stay
low for both, because flowchart-path units were built only for the English NCCN guideline.

\begin{table}[t]
\caption{ElementHit@5 by query type (hybrid retrieval, 199 queries). Best in \textbf{bold}.}
\label{tab:pertype}
\centering
\footnotesize
\setlength{\tabcolsep}{8pt}
\begin{tabular*}{\columnwidth}{@{\extracolsep{\fill}}lccc@{}}
\toprule
\textbf{Method} & \textbf{Table} & \textbf{Flow} & \textbf{Text} \\
\midrule
OCR element          & 0.506 & 0.067 & \textbf{0.535} \\
\textbf{Struct-aware} & \textbf{0.747} & \textbf{0.133} & 0.436 \\
\bottomrule
\end{tabular*}
\end{table}

Splitting by language (114 EN, 85 ZH; Table~\ref{tab:lang}), page-level recall is high in both (PH@5 0.953 ZH vs.\
0.842 EN), so finding the right page is not the bottleneck; the element-level gap (EH@5
0.570 EN vs.\ 0.506 ZH) traces to two construction-coverage issues: flowchart-path units
exist only for the English NCCN guideline (Chinese flowchart queries fall back to coarse
section units, EH@5 $=0$), and the Chinese recommendation regex fires less often (2.7\% of
Chinese units vs.\ 6.6\% of English), diluting Chinese text evidence. Both are addressable
with a multilingual flowchart parser and a Chinese-aware evidence-grade lexicon.

\begin{table}[t]
\caption{Hybrid retrieval by language (EN: 114, ZH: 85 queries).}
\label{tab:lang}
\centering
\footnotesize
\setlength{\tabcolsep}{5pt}
\begin{tabular*}{\columnwidth}{@{\extracolsep{\fill}}llcc@{}}
\toprule
\textbf{Lang.} & \textbf{Method} & \textbf{PH@5} & \textbf{EH@5} \\
\midrule
\multirow{3}{*}{EN}
& Page-level    & 0.842 & N/A \\
& OCR element   & 0.719 & 0.535 \\
& \textbf{Struct-aware} & \textbf{0.886} & \textbf{0.570} \\
\midrule
\multirow{3}{*}{ZH}
& Page-level    & \textbf{0.953} & N/A \\
& OCR element   & \textbf{0.882} & 0.424 \\
& \textbf{Struct-aware} & 0.871 & \textbf{0.506} \\
\bottomrule
\end{tabular*}
\end{table}

\subsection{Context efficiency and ablation}
\label{ssec:efficiency}

Under hybrid retrieval, structure-aware units reach PH@5\,=\,0.879 with 3{,}222 characters
in the top 5 versus page-level's 0.889 with 12{,}111---a statistically indistinguishable
recall ($p{=}0.75$) at 3.8$\times$ less context, and only the structure-aware result
identifies which element is the evidence. Ablating the structural context (searching unit
text alone vs.\ text\,+\,context under BM25) drops PH@5 from 0.809 to 0.769 ($p{=}0.033$)
and MRR$_p$ from 0.658 to 0.603 ($p{<}0.001$), confirming that the attached structural
metadata itself contributes a small but significant gain.

Finally, under a fixed context budget (Table~\ref{tab:budget}), structure-aware units pack
several focused units into tight budgets where a single page-level unit rarely fits: at a
1{,}000-character budget they reach a 0.749 page hit rate versus 0.628 for page-level
retrieval, climbing to 0.889 at 4{,}000 characters while page-level stalls at 0.704.

\begin{table}[t]
\caption{Page hit rate under a fixed context budget (hybrid retrieval).}
\label{tab:budget}
\centering
\footnotesize
\setlength{\tabcolsep}{6pt}
\begin{tabular*}{\columnwidth}{@{\extracolsep{\fill}}lccc@{}}
\toprule
\textbf{Method} & \textbf{1{,}000} & \textbf{2{,}000} & \textbf{4{,}000} \\
\midrule
Flat chunk       & 0.623 & 0.759 & 0.879 \\
Heading chunk    & \textbf{0.764} & \textbf{0.874} & \textbf{0.930} \\
Page-level       & 0.628 & 0.648 & 0.704 \\
OCR element      & 0.658 & 0.769 & 0.879 \\
\textbf{Struct-aware} & 0.749 & 0.819 & 0.889 \\
\bottomrule
\end{tabular*}
\end{table}

\subsection{Error analysis}
\label{ssec:error}

Under hybrid retrieval, structure-aware units place the gold evidence page within the top
5 for 175 of 199 queries (0.879); of the remaining 24, fourteen fall just outside at ranks
6--10 and ten are true misses beyond rank 10. The ten misses skew toward table-dependent
queries (6 of 10) and split 6 English / 4 Chinese. Three cases illustrate the recurring
failure modes. (1)~A table-dependent NCCN query (``systemic therapy options for nonsquamous
vs.\ squamous NSCLC'') never retrieves its page: the answer is split across two algorithm
arms and no single evidence unit covers both---a limitation of per-unit granularity for
comparison-style questions. (2)~A Chinese flowchart-dependent query (short-term insulin
intensification for newly diagnosed T2DM) sinks to rank 67 because flowchart-path units
were built only for the English NCCN guideline; the Chinese decision figure was never
parsed into paths, so the query falls back to a coarse section unit. (3)~An NCCN
text-dependent query (adjuvant chemotherapy indications after R0 resection) reaches rank 22
because the relevant recommendation was absorbed into a longer section unit and diluted.
These point to two concrete improvements: multi-unit evidence aggregation for comparison
queries, and multilingual flowchart parsing so that non-English decision figures also
yield flowchart-path units, both consistent with the language and per-type gaps above.

\section{Discussion and Conclusion}
\label{sec:discussion}

\noindent\textbf{Limitations.} Our unit construction relies on element-type classification
from the parsing stage, so type errors propagate (e.g., a table misclassified as text loses
its row-level units). The flowchart path parsing uses regex-based mermaid extraction that
may miss complex nested structures, and it was built out only for the English NCCN
guideline, so flowchart and Chinese coverage is partial. Two evaluation caveats matter:
element hits are scored by matching the gold element identifier, so page- and chunk-level
baselines that do not expose element boundaries are necessarily N/A rather than zero (the
fair cross-method comparison is therefore at the page level); and our gold annotations were
generated programmatically with manual spot-checking rather than multi-annotator agreement.
The evaluation also stops at retrieval---it does not measure downstream answer correctness
or clinical usability.

\noindent\textbf{Deployment considerations.} The context saving is not merely cosmetic: in
a deployed RAG system, the tokens handed to the LLM dominate both latency and cost, and
oversized retrieval results raise the risk of the answer being drawn from irrelevant
distractor text. Structure-aware units let a system retrieve precisely the table row or
flowchart step that answers a query, which is exactly the granularity at which clinicians
verify an LLM's reasoning against a guideline.

\noindent\textbf{Broader applicability.} Nothing in the construction is specific to
medicine---regulatory standards, technical specifications, and legal codes carry the same
layout-bound relationships, and the same four unit types (or close analogues) apply there.

Within these limits, keeping tables, flowcharts, and graded recommendations intact through
parsing pays off at retrieval time: on 26 guidelines and 199 queries, structure-aware units
consistently rank the gold element highest under BM25, dense, and hybrid
retrieval---significantly so in MRR$_e$---and they match page-level recall at 3.8$\times$
less context. We will release the query set, gold annotations, unit-construction code, and
evaluation scripts to support reproduction and extension to end-to-end answer evaluation.

\section{Compliance with ethical standards}
Only publicly released clinical practice guidelines and public model checkpoints are
used; no human subjects, personal data or private corpora are involved.

\section{Acknowledgements}
No specific funding was received. The query set, gold annotations, unit-construction
code and evaluation scripts will be released upon acceptance. Per the ICASSP 2027 LLM
policy: an AI assistant supported language editing and analysis/plotting code; all
experiments, numbers, code and claims were verified by the authors; no section was
produced wholesale by an LLM.

\bibliographystyle{IEEEbib}
\bibliography{refs}

\end{document}